%% file: Existence2.tex
\documentclass[journal,draftcls,onecolumn,12pt,twoside]{IEEEtran}

\usepackage{cite}

\usepackage[cmex10]{amsmath}
\usepackage{mathtools}
\allowdisplaybreaks[4]

\usepackage{array}
\usepackage{mdwmath}
\usepackage{mdwtab}

\usepackage{url}
\usepackage[normalem]{ulem}
	
\usepackage{graphicx} 
\usepackage{epsfig} 
\usepackage{epstopdf}
\usepackage{amssymb}  
\usepackage{verbatim}
\usepackage[usenames,dvipsnames,svgnames]{xcolor}
\usepackage{multirow} 
\usepackage{dsfont}
\graphicspath{ {./images/} }
\usepackage{verbatim}
\usepackage{subcaption}
\usepackage{wrapfig}
\newcolumntype{L}{>{$}l<{$}}
\usepackage[inline,shortlabels]{enumitem}
\setlist[itemize]{leftmargin=*}
\setlist[enumerate]{leftmargin=*}
\usepackage[hidelinks]{hyperref}	

\newtheorem{theorem}{Theorem}

\newtheorem{lemma}{Lemma}

\DeclareMathOperator*{\argmax}{argmax}

\newcommand{\mX}{\mathcal{X}}

\newcommand{\mN}{\mathcal{N}}

\renewcommand{\hat}{\widehat}

\newcommand{\uF}{\underline{F}}
\newcommand{\mP}{\mathbb{P}}

\newcommand{\eq}[1]{\begin{align}#1\end{align}}
\newcommand{\seq}[1]{\begin{subequations}#1\end{subequations}}

\newcommand{\E}{\mathbb{E}}
 \newcommand{\nn}{\nonumber}
\newcommand{\cX}{\mathcal{X}}
\newcommand{\cA}{\mathcal{A}}

\newcommand{\cN}{\mathcal{N}}

\newcommand{\defeq}{\buildrel\triangle\over =}

\newcommand{\pushright}[1]{\ifmeasuring@ #1 \else\omit\hfill$\displaystyle#1$\fi\ignorespaces}
\newcommand{\pushleft}[1]{\ifmeasuring@ #1 \else\omit$\displaystyle#1$\hfill\fi\ignorespaces}
\newcommand{\hs}[1]{\hspace{#1}}
\newcommand{\lp}{\left.}
\newcommand{\rp}{\right.}
\begin{document}
	%
	\title{Existence of structured perfect Bayesian equilibrium in dynamic games of asymmetric information}
	\author{Deepanshu Vasal\thanks{email:dvasal@umich.edu}}
	%
	%
	%
	
	
	
	%
	\maketitle
	\begin{abstract}
	In~\cite{VaSiAn19}, authors considered a general finite horizon model of dynamic game of asymmetric information, where $N$ players have types evolving as independent Markovian process, where each player observes its own type perfectly and actions of all players. The authors present a sequential decomposition algorithm to find all \textit{structured perfect Bayesian equilibria} of the game. The algorithm consists of solving a class of fixed-point of equations for each time $t$, whose existence was left as an open question. In this paper, we prove existence of these fixed-point equations for compact metric spaces.
	\end{abstract}
	\section{Introduction}
	Authors in~\cite{VaSiAn19} considered a model consisting of strategic players having types that evolve as conditionally independent Markov controlled processes. Players observe their types privately and actions taken by all players are observed publicly. Instantaneous reward for each player depends on everyone's types and actions. The proposed methodology provides a decomposition of the interdependence between beliefs and strategies in PBE and enables a systematic evaluation of a subset of PBE, namely \textit{structured perfect Bayesian equilibria} (SPBE). 
Here SPBE are defined as equilibria with players strategies based on their current private type and a set of beliefs on each player's current private type, which is common to all the players and whose domain is time-invariant. The beliefs on players' types are such that they can be updated individually for each player and sequentially w.r.t. time. The model allows for signaling amongst players as beliefs depend on strategies.
They present a sequential decomposition methodology for calculating all SPBEs for finite and infinite horizon dynamic games with asymmetric information.

For the finite horizon model, they provide a two-step algorithm involving a backward recursion followed by a forward recursion. The algorithm works as follows. In the backward recursion, for every time period, the algorithm finds an equilibrium generating function defined for all possible common beliefs at that time. This involves solving a one-step fixed point equation on the space of probability simplexes. Existence of this fixed-point equation was left as an open question. If there exists a solution for every common belief $\pi_t$, then the equilibrium strategies and beliefs are obtained through a forward recursion using the equilibrium generating function obtained in the backward step and the Bayes update rule. 
A similar notion of equilibrium was defined in~\cite{OuTaTe15,Ta17} where authors provide a sequential decomposition algorithm to compute common information based perfect Bayesian equilibrium. In~\cite{Ta17}, it was shown that such an equilibrium exists for zero-sum games.

In this paper, we consider the finite horizon game with all sets of variables in a compact metric space. We show that there always exists an SPBE for such a game. Since it is proved in~\cite{VaSiAn19} that all SPBEs can be found using their algorithm, we conclude that there always exists a solution to the fixed-point equation considered in~\cite{VaSiAn19} for each time period $t$. 
	\section{Model and Preliminaries}
\label{sec:Model}
We consider a discrete-time dynamical system with $N$ strategic players in the set $ \cN \defeq \{1,2, \ldots N \}$. {We consider two cases: finite horizon $\mathcal{T} \defeq \{1, 2, \ldots T\}$ with perfect recall and infinite horizon with perfect recall}. The system state is $X_t \defeq (X_t^1, X_t^2, \ldots X_t^N)$, where $X_t^i \in \cX^i$ is the type of player $i$ at time $t$, which is perfectly observed and is its private information. Players' types evolve as conditionally independent, controlled Markov processes such that
\seq{
\eq{
\mP(x_1) &= \prod_{i=1}^N Q^i_1(x_1^i)\\
\mP(x_t|x_{1:t-1}, a_{1:t-1}) &= \mP(x_t|x_{t-1} , a_{t-1})\\
&= \prod_{i=1}^N Q_t^{i}(x_t^i|x_{t-1}^i, a_{t-1}),
}
}
where $Q^i_t$ are known kernels. Player $i$ at time $t$ takes action $a_t^i \in \cA^i$ on observing the actions {$a_{1:t-1} = (a_k)_{k=1,\ldots,t-1}$ where $ a_k = \left( a_k^j \right)_{j \in \mN} $}, which is common information among players, and the types $x_{1:t}^i$, which it observes privately. The sets $\cA^i, \cX^i $ are assumed to be finite. Let $g^i = ( g^i_t)_{t \in \mathcal{T}}$ be a probabilistic strategy of player $i$  where $g^i_t : \cA^{t-1}\times (\cX^{i})^t \to \Delta(\cA^i)$ such that player $i$ plays action $A_t^i$ according to $ A_t^i \sim g^i_t(\cdot|a_{1:t-1},x_{1:t}^i)$. Let $ g \defeq(g^i)_{i\in \cN}$ be a strategy profile of all players. At the end of interval $t$, player $i$ receives an instantaneous reward $R_t^i(x_t,a_t)$. {To preserve the information structure of the problem, it is assumed that players do not observe their rewards until the end of game.\footnote{Alternatively, we could have assumed instantaneous reward of a player to depend only on its own type, i.e. be of the form $R_t^i(x_t^i,a_t)$, and have allowed rewards to be observed by the players during the game as this would not alter the information structure of the game} The reward functions and state transition kernels are common knowledge among the players.} For the finite-horizon problem, the objective of player $i$ is to maximize its total expected reward
\eq{ J^{i,g} \defeq \E^g \left\{ \sum_{t=1}^T R_t^i(X_t,A_t) \right\}.
}

\subsection{Preliminaries}
\label{sec:Result_A}
Any history of this game at which players take action is of the form $h_t = (a_{1:t-1},x_{1:t})$. Let $\mathcal{H}_t$ be the set of such histories, $\mathcal{H}^T \defeq \cup_{t=0}^T \mathcal{H}_t $ be the set of all possible such histories in finite horizon and $\mathcal{H}^\infty \defeq \cup_{t=0}^\infty \mathcal{H}_t $ for infinite horizon. At any time $t$ player $i$ observes $h^i_t = (a_{1:t-1},x_{1:t}^{i})$ and all players together have $h^c_t = a_{1:t-1}$ as common history. Let $\mathcal{H}^i_t$ be the set of observed histories of player $i$ at time $t$ and $\mathcal{H}^c_t$ be the set of common histories at time $t$. An appropriate concept of equilibrium for such games is PBE \cite{FuTi91book}, which consists of a pair $(\beta^*,\mu^*)$ of strategy profile $\beta^* = (\beta_t^{*,i})_{t \in \mathcal{T},i\in \cN}$ where $\beta_t^{*,i} : \mathcal{H}_t^i \to \Delta(\cA^i)$ and a belief profile $\mu^* = (^i\mu_t^{*})_{t \in \mathcal{T},i\in \cN}$ where $^i\mu_t^{*}: \mathcal{H}^i_t \to \Delta(\mathcal{H}_t)$ that satisfy sequential rationality so that $\forall i \in \cN,  t \in \mathcal{T},  h^{i}_t \in \mathcal{H}^i_t, {\beta^{i}}$
\begin{equation}
W_t^{i,\beta^{*,i},T}(h_t^i) \ge W_t^{i,\beta^i,T}(h_t^i)
\end{equation}
where the reward-to-go  is defined as
\begin{equation}
\hs{-0.2cm}W_t^{i,\beta^i,T}(h_t^i) \triangleq \E^{{\beta}^{i} \beta^{*,-i},\, ^i\mu_t^*[h_t^i]}\left\{ \sum_{n=t}^T R_n^i(X_n, A_n)\big\lvert  h^i_t\right\}, \;\;   \label{eq:seqeq}
\end{equation}
and the beliefs satisfy some consistency conditions as described in~\cite[p. 331]{FuTi91book}. 
In general, a belief for player $i$ at time $t$, $^i\mu_t^{*}$ is defined on history $h_t = (a_{1:t-1},x_{1:t}) $ given its private history $h^i_t = (a_{1:t-1},x_{1:t}^{i})$. Here player $i$'s private history $h^i_t=(a_{1:t-1},x_{1:t}^i)$ consists of a public part $h_t^c=a_{1:t-1}$ and a private part $x_{1:t}^i$.
At any time $t$, the relevant uncertainty player $i$ has is about other players' types $x_{1:t}^{-i} \in \times_{n=1}^t \left( \times_{j \ne i} \mX^j \right)$  and their future actions.
Due to independence of types, and given the common history $h_t^c$, player $i$'s type history $x_{1:t}^i$ does not provide any additional information about $x_{1:t}^{-i}$, as will be shown later. For this reason only those beliefs are considered that are functions of each player's history $h^i_t$ only through the common history $h^c_t$.
Hence, for each player $i$, its belief for each history $h^c_t=a_{1:t-1}$ is derived from a common belief $\mu^*_t[a_{1:t-1}]$. Furthermore, as will be shown later, this belief factorizes into a product of marginals $\prod_{j\in\cN} \mu^{*,j}_t[a_{1:t-1}] $.
Thus the following system of beliefs can suffiently be used, $\mu^*=(\underline{\mu}^*_t)_{t\in\mathcal{T}}$, where  $\underline{\mu}^*_t = (\mu^{*,i}_t)_{i \in \mN} $, and $\mu^{*,i}_t: \mathcal{H}^c_t \to \Delta(\cX^i)$, with the understanding that player $i$'s belief on $x_t^{-i}$ is  $\mu^{*,-i}_t[a_{1:t-1}](x_t^{-i})=\prod_{j\neq i} \mu^{*,j}_t[a_{1:t-1}](x_t^j)$.
Under the above structure, all consistency conditions that are required for PBEs~\cite[p. 331]{FuTi91book} are automatically satisfied.
	
\section{Structured perfect Bayesian algorithm}

At any time $t$, player $i$ has information $(a_{1:t-1}, x_{1:t}^{i})$ where $a_{1:t-1}$ is the common information among players, and $x_{1:t}^{i}$ is the private information of player $i$. Since $(a_{1:t-1}, x_{1:t}^{i})$ increases with time, any strategy of the form $A_t^i \sim g^i_t(\cdot|a_{1:t-1},x_{1:t}^{i})$ becomes unwieldy. Thus it is desirable to have an information state in a time-invariant space that succinctly summarizes $(a_{1:t-1}, x_{1:t}^{i})$, and that can be sequentially updated. I
For any strategy profile $g$, belief $\pi_t $ on $X_t$, $\pi_t \in \Delta (\cX)$, is defined as $\pi_t (x_t) \defeq \mP^g(X_t=x_t|a_{1:t-1}), \; \forall x_t\in \cX$. Define the marginals $\pi_t^{{i}} (x_t^i) \defeq \mP^g(X_t^i=x_t^i|a_{1:t-1}), \; \forall x_t^i \in \cX^i$.

 Then using common information approach~\cite{NaMaTe13}, the problem is posed as follows: player $i$ at time $t$ observes $a_{1:t-1}$ and takes action $\gamma_t^i$, where $\gamma_t^i :  \cX^i \to \Delta(\cA^i)$ is a partial (stochastic) function from its private information $x_t^i$ to  $a_t^i$, of the form $A_t^i \sim \gamma_t^i(\cdot|x_t^i)$. These actions are generated through some policy $\psi^i = (\psi^i_t)_{t \in \mathcal{T}}$, $\psi^i_t : \cA^{t-1} \to \left\{  \cX^i \to \Delta(\cA^i) \right\}$, that operates on the common information $a_{1:t-1}$ such that $\gamma_t^i = \psi_t^i[a_{1:t-1}]$. Then any policy of the form $A_t^i \sim s_t^i(\cdot|a_{1:t-1},x_t^{i})$ is equivalent to $A_t^i \sim \psi^i_t[a_{1:t-1}] (\cdot|x_t^i)$.

\subsection{Forward/Backward algorithm}
In Lemma~3 in Appendix~B of~\cite{VaSiAn19}, it is shown that due to the independence of types and their evolution as independent controlled Markov processes, for any strategy of the players, the joint common belief can be factorized as a product of its marginals i.e., $\pi_t(x_t) = \prod_{i=1}^N \pi_t^{i}(x_t^i), \forall x_t$. Define $\underline{\pi}_t \in \times_{i\in \cN} \Delta(\cX^i)$ as vector of marginal beliefs where $\underline{\pi}_t := (\pi^i_t)_{i\in \cN}$. Similarly define the vector of belief updates as $\underline{F}(\underline{\pi},\gamma,a) := (F^i(\pi^i,\gamma^i,a))_{i \in \cN}$ where (using Bayes rule)
\eq{\label{eq:F_update}
&F^i(\pi^i,\gamma^i,a)(x_{t+1}^i)= \nn \\
& \left\{
\begin{array}{ll}
\frac{  \sum_{x^i_t} \pi^{{i}}(x_t^i) \gamma^i(a^i|x_t^i)Q_t^i(x_{t+1}^i|x_t^i, a) }{ \sum_{\tilde{x}_t^i}\pi^{{i}}(\tilde{x}_t^i)  \gamma^i(a^i|\tilde{x}_t^i)}   \mbox{if } \sum_{\tilde{x}_t^i}\pi^{{i}}(\tilde{x}_t^i)  \gamma^i(a^i|\tilde{x}_t^i) > 0 \\
\sum_{x_t^i}\pi^i(x_t^i)Q_t^i(x_{t+1}^i|x_t^i,a) \;\;\;\;\; \mbox{if } \sum_{\tilde{x}_t^i}\pi^{{i}}(\tilde{x}_t^i)  \gamma^i(a^i|\tilde{x}_t^i) = 0.
\end{array}
\right.
}

In the following, a backward-forward algorithm is presented that evaluates SPBE. where it is shown in \cite[Theorem~2]{VaSiAn19}, this is a ``canonical'' methodology, in the sense that all SPBE can be generated this way.

\subsection{Backward Recursion} \label{sec:fhbr}

In this section, define an equilibrium generating function \\$\theta=(\theta^i_t)_{i\in\cN,t\in\mathcal{T}}$, where $\theta^i_t : \times_{i\in\cN} \Delta(\cX^i) \to \left\{\cX^i \to \Delta(\cA^i) \right\}$. In addition, define a sequence of reward-to-go functions of player $i$ at time $t$,  $(V_t^i)_{i\in \cN, t\in \{ 1,2, \ldots T+1\}}$, where $V_t^i : \times_{i\in\cN} \Delta(\cX^i) \times \cX^i \to \mathbb{R}$.
These quantities are generated through a backward recursive way, as follows.
\begin{itemize}
\item[1.] Initialize $\forall \underline{\pi}_{T+1}\in \times_{i\in\cN} \Delta(\cX^i), x_{T+1}^i\in \cX^i$,
\eq{
V^i_{T+1}(\underline{\pi}_{T+1},x_{T+1}^i) \defeq 0.   \label{eq:VT+1}
}

\item[2.] For $t = T,T-1, \ldots 1, \ \forall \underline{\pi}_t \in \times_{i\in\cN} \Delta(\cX^i), \pi_t = \prod_{i\in\cN}\pi_t^i $, let $\theta_t[\underline{\pi}_t] $ be generated as follows. Set $\tilde{\gamma}_t = \theta_t[\underline{\pi}_t]$, where $\tilde{\gamma}_t$ is the solution, if it exists, of the following fixed-point equation, $\forall i \in \cN,x_t^i\in \cX^i$,
  \eq{
 \tilde{\gamma}^{i}_t(\cdot|x_t^i) \in \arg\max_{\gamma^i_t(\cdot|x_t^i)} &\E^{\gamma^i_t(\cdot|x_t^i) \tilde{\gamma}^{-i}_t,\,\pi_t} \left\{ R_t^i(X_t,A_t) + \rp\nn\\
 &\hs{-0.2cm}\lp V_{t+1}^i (\uF(\underline{\pi}_t, \tilde{\gamma}_t, A_t), X_{t+1}^i) \big\lvert x_t^i \right\} , \label{eq:m_FP}
  }
 where expectation in \eqref{eq:m_FP} is with respect to random variables $(X_t^{-i},A_t, X_{t+1}^i)$ through the measure
$\pi_t^{-i}(x_t^{-i})\gamma^i_t(a^i_t|x_t^i) \tilde{\gamma}^{-i}_t(a^{-i}_t|x_t^{-i})Q_{t+1}^i(x_{t+1}^i|x_t^i,a_t)$ and $\uF$ is defined above.

 Furthermore, using the quantity $\tilde{\gamma}_t$ found above, define
  \eq{
  V^i_{t}(\underline{\pi}_t,x_t^i) \defeq & \E^{\tilde{\gamma}^{i}_t(\cdot|x_t^i) \tilde{\gamma}^{-i}_t,\, \pi_t}\left\{ {R}_t^i (X_t,A_t) \rp\nn\\
  &\lp+ V_{t+1}^i (\uF(\underline{\pi}_t, \tilde{\gamma}_t, A_t), X_{t+1}^i)\big\lvert  x_t^i \right\}.  \label{eq:Vdef}
   }
   \end{itemize}
   
   Then
   
   \begin{theorem}[\cite{VaSiAn19}]
\label{Thm:Main}
A strategy and belief profile $(\beta^*,\mu^*)$, constructed through the backward-forward recursion algorithm is a PBE of the game, i.e.,
$\forall i \in \cN,t \in \mathcal{T}, a_{1:t-1} \in \mathcal{H}_t^c, x_{1:t}^i \in (\cX^i)^t, \beta^i$,
\eq{
&\E^{\beta_{t:T}^{*,i} \beta_{t:T}^{*,-i},\,\mu_{t}^{*}[a_{1:t-1}]} \left\{ \sum_{n=t}^T R_n^i(X_n,A_n) \big\lvert  a_{1:t-1}, x_{1:t}^i \right\} \geq\nn\\
&\E^{\beta_{t:T}^{i} \beta_{t:T}^{*,-i},\, \mu_{t}^{*}[a_{1:t-1}]} \left\{ \sum_{n=t}^T R_n^i(X_n,A_n) \big\lvert  a_{1:t-1}, x_{1:t}^i \right\}. \label{eq:prop}
}
\end{theorem}

   \begin{theorem}[\cite{VaSiAn19}]
	\label{thm:2}
	Let ($\beta^*,\mu^*$) be an SPBE. Then there exists an equilibrium generating function $\phi$
 that satisfies \eqref{eq:m_FP} in backward recursion $\forall\ \pi_t = \mu^*_t[a_{1:t-1}], \ \forall\ a_{1:t-1}$,
	such that  ($\beta^*,\mu^*$) is defined through forward recursion using $\phi$.
\end{theorem}
	
\section{Main result: Existence}
In this section, we present the main result of this paper where we show that for all $\forall \underline{\pi}_t \in \times_{i\in\cN} \Delta(\cX^i), \pi_t = \prod_{i\in\cN}\pi_t^i$, there always exists a $\tilde{\gamma} = \theta[\pi_t]$ that satisfies~\eqref{eq:m_FP}. We show that existence holds true for any compact metric spaces $\cX,\cA$.
Let $\Sigma$ be the set of all mixed strategies of the form $\sigma_t^i(\cdot|a_{1:t-1},x_{1:t}^i)$ and $\Sigma'$ be the set of all mixed strategies of the form $\sigma_t^i(\cdot|a_{1:t-1},x_{t}^i)$. Then both $\Sigma,\Sigma'$ are compact.
\begin{lemma}
Let $\epsilon>0$.

\begin{itemize}
\item[(a)] Let $L$ be a standard measure on $\mathcal{A}^i$ such that for any non empty open measurable set $V$ in the Borel space $B(\mathcal{A}^i)$, $L(V)>0$ i.e. L puts non zero measure on every non-empty open measurable set of $\mathcal{A}^i$.\footnote{L can be Lesbesgue measure on Euclidean space and counting measure on a discrete space.} Let $\Sigma^{\epsilon}_1\subset \Sigma'$ be the space of such strategies such that if $\sigma\in\Sigma^{\epsilon}_1$ then for all $i\in\mathcal{N},V\in B(\mathcal{A}^i),a_{1:t-1},x_{t}^i$,
\eq{
\sigma^i(V|a_{1:t-1},x_{t}^i)\geq \epsilon L(V)
}  

If $\mathcal{A}^i$ is discrete the above condition boils down to: for all $i\in\mathcal{N},a^j\in\mathcal{A}^i,a_{1:t-1},x_{t}^i$,
\eq{
\sigma^i(a^j|a_{1:t-1},x_{t}^i)\geq \epsilon 
}  

\item[(b)] Let $\Sigma^{\epsilon}_2\subseteq \Sigma^{\epsilon}_1$ be the space of such strategies such that if $\sigma\in\Sigma^{\epsilon}_2$ then for all $a_{1:t-1},\hat{a}_{1:t-1}$ such that if
\eq{
P^{\sigma}(x_t|a_{1:t-1}) = P^{\sigma}(x_t|\hat{a}_{1:t-1}) \;\;\;\;\; \forall x_t,
}
it implies,
\eq{
\sigma_t^i(\cdot|a_{1:t-1},x_t^i) =  \sigma_t^i(\cdot|\hat{a}_{1:t-1},x_t^i) \;\;\;\; \forall i, x_t^i.
}
(The above condition ensures the property of structured policies such that if two action histories produce the same common belief under a strategy, then the strategy is indifferent towards those two action histories). 
 \end{itemize}
 Then $\Sigma^{\epsilon}_2$ is a compact metric space.
 \end{lemma}
 \begin{IEEEproof}
 \begin{itemize}
 \item[(a)] Let $\sigma^n$ be any sequence of strategies in $\Sigma^{\epsilon}_1$. Thus $i\in\mathcal{N},V\in B(\mathcal{A}^i),a^j,a_{1:t-1},x_{t}^i$, \\$\sigma^{i,n}(V|a_{1:t-1},x_{t}^i)\geq \epsilon L(V)$. Thus $\lim_n \sigma^{i,n}(V|a_{1:t-1},x_{t}^i)\geq \epsilon L(V)$. Thus $\Sigma^{\epsilon}_1$ is closed. It is also bounded.  
 Thus $\Sigma^{\epsilon}_1$ is compact.
 
 \item[(b)] Let $\mathcal{S}(a_{1:t-1},\hat{a}_{1:t-1}):= \{\sigma\in\Sigma^{\epsilon}_1:\forall x_t,\; P^{\sigma}(x_t|a_{1:t-1}) = P^{\sigma}(x_t|\hat{a}_{1:t-1})\}$. Clearly $\mathcal{S}(a_{1:t-1},\hat{a}_{1:t-1})$ is not empty since the strategy that is independent of action history always lies in this set.
 
 Let $\mathcal{R}(a_{1:t-1},\hat{a}_{1:t-1}):=\{\sigma\in\Sigma^{\epsilon}_1: \forall i,x_t^i\;\; \sigma_t^i(\cdot|a_{1:t-1},x_t^i) =  \sigma_t^i(\cdot|\hat{a}_{1:t-1},x_t^i)\}$. $\mathcal{R}(a_{1:t-1},\hat{a}_{1:t-1})$ is non empty for the same reason.
 
 Then $\Sigma^{\epsilon}_2 = \bigcap_{a_{1:t-1},\hat{a}_{1:t-1}}\big(\mathcal{S}^c(a_{1:t-1},\hat{a}_{1:t-1})\bigcup \mathcal{R}(a_{1:t-1},\hat{a}_{1:t-1})\big)$.
 

 Let $\mathcal{V}(a_{1:t-1},\hat{a}_{1:t-1}):=\{g\in \mathbb{R}^{|\mathcal{X}|\times|\mathcal{A}|^{t-1}}: \;\; g(a_{1:t-1}) =  g(\hat{a}_{1:t-1})$\}. Since $\mathcal{S}(a_{1:t-1},\hat{a}_{1:t-1})$ is intersection of ${\Sigma}^{\epsilon}_1$, which is compact, with subspace $\mathcal{V}(a_{1:t-1},\hat{a}_{1:t-1})$, thus $\mathcal{S}(a_{1:t-1},\hat{a}_{1:t-1})$ is also compact.

 Let $\mathcal{W}(a_{1:t-1},\hat{a}_{1:t-1}):=\{ f=(f^i)_i, f^i\in \mathbb{R}^{|\mathcal{X}|\times |\mathcal{A}|^{t-1}\times |\mathcal{X}^i|}:\forall i,x_t^i,\;\; f^i(a_{1:t-1},x_t^i) =  f^i(\hat{a}_{1:t-1},x_t^i)$\}.

Then $\mathcal{R}(a_{1:t-1},\hat{a}_{1:t-1})$ is the intersection of $\mathcal{S}(a_{1:t-1},\hat{a}_{1:t-1})$ with subspace $\mathcal{W}(a_{1:t-1},\hat{a}_{1:t-1})$, thus $\mathcal{R}(a_{1:t-1},\hat{a}_{1:t-1})$ is also compact.

This implies $\Sigma^{\epsilon}_2$ is non-empty and compact. 

%
%
%
 \end{itemize}
 
 \end{IEEEproof}
 
\begin{theorem}
There exists an SPBE for such a game.
\end{theorem}
\begin{IEEEproof}
 Define a perturbation of the game by restricting the set of strategies to $\Sigma^{\epsilon}$.
Since utilities are continuous in strategies on $\Sigma^{\epsilon}$ (as all strategies put non zero measure on every non empty open set in the action space which implies the Bayes rule has denominator non-zero), and $\Sigma^{\epsilon}$ is a compact metric space by Lemma~1.
Let 
\eq{
BR^i(\beta^{-i}) = \bigcap_{\beta^i \in\Sigma_2^{\epsilon,i}}\argmax_{\beta^i \in\Sigma^{\epsilon,i}} \E^{\beta^{i} \beta^{-i}} \left\{ \sum_{n=1}^T R_n^i(X_n,A_n) \right\}.\label{eq:2}
}
We note that $BR^i(\beta^{-i})$ is non-empty $\forall \beta^{-i}$ by Lemma~2 in~\cite{VaSiAn19} which states that any expected reward that can be achieved using any general profile of strategies, can also be achieved using structured policies, which implies when the intersection of the argmax operation in~\eqref{eq:2} is taken over structured policies, it is non-empty. Moreover $BR^i(\beta^{-i})$ is continuous in $\beta^{-i}$ by~\cite[Theorem~2,3]{VaBe20}.

Since $BR$ is continuous on $\Sigma^{\epsilon}$, there exists a fixed-point of $BR$ by Glicksberg fixed-point Theorem~\cite{Gl52}. Consider a sequence of such perturbed games in which  $\epsilon \to 0$; by the compactness of the set of strategy profiles, some sequence of selections from the sets of Nash equilibria of the games in the sequence converges, say to $\sigma^*$. Then $\sigma^*$ is an SPBE of the game.
\end{IEEEproof}

\begin{theorem}
There exists a solution to the fixed-point equation~\eqref{eq:m_FP} for each $t,\pi_t$.
\end{theorem}
\begin{IEEEproof}
Since there exists an SPBE for such a game from Theorem~3 and every SPBE can be found from~\cite[Theorem~4]{VaSiAn19}, thus there exists a solution to~\eqref{eq:m_FP} for each $t,\pi_t$.
\end{IEEEproof}

\textbf{Remark}
In this paper, we showed that there exists a solution of smaller fixed-point equations~\eqref{eq:m_FP} for each $t,\pi_t$ by showing that there exists a solution to the fixed-point equation which is the Nash equilibrium of the whole game, using Glicksberg Fixed-point equation. It is interesting because the smaller fixed-point equations may and does have discontinuous utilities, as seen through numerical solutions. There have been some results known in existence of solutions of games with discontinuous utilities~\cite{DaMa86i,DaMa86ii}. It would be an interesting exercise to explore the connection between existence of the two results.

\section{Conclusion}
Authors in~\cite{VaSiAn19} proposed a novel methodology to find structured perfect Bayesian equilibria of a general model of dynamic games of asymmetric information where players observe their types privately, which evolve as a controlled Markov process, condition on everyones actions. Players also observe everyone's actions publicly. In this paper, we prove the existence of the fixed-point equation that crucially appears in their methodology for each time $t$ and that was left open in that paper.

\section{Acknowledgement}
The author would like to thank K.S. Mallikarjuna Rao for suggestion to prove existence of SPBE.
\bibliographystyle{IEEEtran}
\input{root.bbl}

\end{document}

%% file: root.bbl

%% file: Existence2.bbl
\begin{thebibliography}{}
\providecommand{\url}[1]{#1}
\csname url@samestyle\endcsname
\providecommand{\newblock}{\relax}
\providecommand{\bibinfo}[2]{#2}
\providecommand{\BIBentrySTDinterwordspacing}{\spaceskip=0pt\relax}
\providecommand{\BIBentryALTinterwordstretchfactor}{4}
\providecommand{\BIBentryALTinterwordspacing}{\spaceskip=\fontdimen2\font plus
\BIBentryALTinterwordstretchfactor\fontdimen3\font minus
  \fontdimen4\font\relax}
\providecommand{\BIBforeignlanguage}[2]{{%
\expandafter\ifx\csname l@#1\endcsname\relax
\typeout{** WARNING: IEEEtran.bst: No hyphenation pattern has been}%
\typeout{** loaded for the language `#1'. Using the pattern for}%
\typeout{** the default language instead.}%
\else
\language=\csname l@#1\endcsname
\fi
#2}}
\providecommand{\BIBdecl}{\relax}
\BIBdecl

\end{thebibliography}


\begin{thebibliography}{1}
\providecommand{\url}[1]{#1}
\csname url@samestyle\endcsname
\providecommand{\newblock}{\relax}
\providecommand{\bibinfo}[2]{#2}
\providecommand{\BIBentrySTDinterwordspacing}{\spaceskip=0pt\relax}
\providecommand{\BIBentryALTinterwordstretchfactor}{4}
\providecommand{\BIBentryALTinterwordspacing}{\spaceskip=\fontdimen2\font plus
\BIBentryALTinterwordstretchfactor\fontdimen3\font minus
  \fontdimen4\font\relax}
\providecommand{\BIBforeignlanguage}[2]{{%
\expandafter\ifx\csname l@#1\endcsname\relax
\typeout{** WARNING: IEEEtran.bst: No hyphenation pattern has been}%
\typeout{** loaded for the language `#1'. Using the pattern for}%
\typeout{** the default language instead.}%
\else
\language=\csname l@#1\endcsname
\fi
#2}}
\providecommand{\BIBdecl}{\relax}
\BIBdecl

\bibitem{VaSiAn19}
D.~Vasal, A.~Sinha, and A.~Anastasopoulos, ``{A Systematic Process for
  Evaluating Structured Perfect Bayesian Equilibria in Dynamic Games With
  Asymmetric Information},'' \emph{IEEE Transactions on Automatic Control},
  vol.~64, no.~1, pp. 81--96, jan 2019.

\bibitem{OuTaTe15}
Y.~Ouyang, H.~Tavafoghi, and D.~Teneketzis, ``{Dynamic oligopoly games with
  private {\{}M{\}}arkovian dynamics},'' in \emph{Proc. 54th IEEE Conf.
  Decision and Control (CDC)}, 2015.

\bibitem{Ta17}
H.~T. Jahormi, ``{On Design and Analysis of Cyber-Physical Systems with
  Strategic Agents},'' Ph.D. dissertation, University of Michigan, Ann Arbor,
  2017.

\bibitem{FuTi91book}
D.~Fudenberg and J.~Tirole, \emph{{Game Theory}}.\hskip 1em plus 0.5em minus
  0.4em\relax Cambridge, MA: MIT Press, 1991.

\bibitem{NaMaTe13}
A.~Nayyar, A.~Mahajan, and D.~Teneketzis, ``Decentralized stochastic control
  with partial history sharing: A common information approach,''
  \emph{Automatic Control, IEEE Transactions on}, vol.~58, no.~7, pp.
  1644--1658, 2013.

\bibitem{VaBe20}
\BIBentryALTinterwordspacing
D.~Vasal and R.~Berry, ``{\$}alpha-{\$} robust equilibrium in anonymous
  games,'' may 2020. [Online]. Available: \url{http://arxiv.org/abs/2005.06812}
\BIBentrySTDinterwordspacing

\bibitem{Gl52}
I.~L. Glicksberg, ``{A Further Generalization of the Kakutani Fixed Point
  Theorem, with Application to Nash Equilibrium Points},'' \emph{Proceedings of
  the American Mathematical Society}, vol.~3, no.~1, p. 170, feb 1952.

\bibitem{DaMa86i}
P.~Dasgupta, E.~Maskin, and Others, ``{The existence of equilibrium in
  discontinuous economic games, I: Theory},'' \emph{Review of Economic
  Studies}, vol.~53, no.~1, pp. 1--26, 1986.

\bibitem{DaMa86ii}
P.~Dasgupta and E.~Maskin, ``{The existence of equilibrium in discontinuous
  economic games, II: Applications},'' \emph{The Review of economic studies},
  vol.~53, no.~1, pp. 27--41, 1986.

\end{thebibliography}
